\documentclass{aa}
\usepackage{txfonts}
\usepackage{graphicx}
\usepackage{natbib}

\begin{document}

\title{Highly Ionized Fe X-ray Lines at Energies 7.7--8.6 keV}

\author{K. J. H. Phillips\inst{1}}

\institute{UCL--Mullard Space Science Laboratory, Holmbury St Mary, Dorking, Surrey RH5 6NT, United Kingdom\\
\email{kjhp@mssl.ucl.ac.uk} }

\offprints{K. J. H. Phillips \email{kjhp@mssl.ucl.ac.uk}}

\date{Received   /Accepted     }

\abstract {\ion{Fe}{xxv} lines at 1.85~\AA\ (6.70~keV) and nearby \ion{Fe}{xxiv}
satellites have been widely used for determining the temperature of the hottest parts of
solar flare and tokamak plasmas, though the spectral region is crowded and the lines are
blended during flare impulsive stages.}{The aim of this work is to show that similarly
excited Fe lines in the 7.7--8.6~keV (1.44--1.61~\AA) region have the same diagnostic
capability with the advantage of not being so crowded. } {Spectra in the 7.7--8.6~keV
range are synthesized using the CHIANTI spectral package for conditions (temperature,
turbulent velocities) appropriate to solar flares. } {The calculated spectra show that
the Fe lines in the 7.7--8.6~keV are well separated even when turbulent velocities are
present, and \ion{Fe}{xxiv}/\ion{Fe}{xxv} line ratios should therefore provide valuable
tools for diagnosing flares and tokamak plasmas. }{Fe lines in the 7.7--8.6~keV range are
ideal for the measurement of flare temperature and for detecting the presence of
low-energy nonthermal electrons present at flare impulsive stages. An indication of what
type of instruments to observe this region is given.}


\maketitle

\section{Introduction}

The diagnostic potential of highly ionized iron lines at $\sim 1.9$~\AA\ ($\sim 6.7$~keV)
was recognized by \cite{gab72} who pointed out that \ion{Fe}{xxiv} satellite lines with
transitions $1s^2 nl - 1s 2p nl$ ($n\geqslant 2$), with upper levels formed by
dielectronic recombination of He-like iron (Fe$^{+24}$), have fluxes which, relative to
the \ion{Fe}{xxv} resonance line $w$ ($1s^2\,^1S_0 - 1s2p\,^1P_1$) at 1.851~\AA\
(6.699~keV), depend on electron temperature $T_e$, roughly as $T_e^{-1}$. This has proved
extremely useful for understanding the hottest parts of solar flare plasmas \citep{dos90}
and tokamak plasmas \citep{bit79}. The most intense $n=2$ satellites include (in the
notation of \cite{gab72}) $j$ (1.867~\AA, 6.642~keV) and $k$ (1.864~\AA, 6.652~keV) which
are resolved from other nearby lines by crystal spectrometers with spectral resolution
0.001~\AA\ (4~eV). The most intense $n=3$ satellites include the $d13$, $d15$ pair at
1.852~\AA\ (6.695~keV), which are on the long-wavelength side of line $w$ \citep{bel79a},
while $n\gtrsim 4$ satellites are progressively less separated from the $w$ line
\citep{bel79b}. At the onset of solar flares, X-ray lines show a broadening which is
commonly attributed to turbulence resulting from the impact of nonthermal electrons
accelerated at the flare impulsive stage. The broadening may amount to 0.03~\AA\
(0.11~keV, equivalent to a few hundred km~s$^{-1}$) which leads to a blurring of the
satellite line structure, so determinations of temperatures become less certain. Also,
satellites with $n\geqslant 3$ on the long-wavelength side of the \ion{Fe}{xxv} line $w$
including the $d13$, $d15$ pair may be indistinguishable from $w$. A method for searching
for low-energy, nonthermal electrons proposed by \cite{gab79}, involving the comparison
of temperatures derived from the ratios $j/w$ and $(d13+d15)/w$, is thus made difficult
to apply. In addition to the turbulent broadening, a short-wavelength (``blue-shifted")
component is generally present for disk flares, making the determination of temperatures
from line flux ratios even less certain.

While the spectral region 1.851--1.940~\AA\ (6.391--6.699~keV), which includes all
dielectronic satellites emitted by \ion{Fe}{xxiv} and lower ionization stages of Fe, has
been very well studied with high-resolution spectrometers, the region 1.44--1.60~\AA\
(7.7--8.6~keV) has hardly received any attention. This range contains higher-$n$
\ion{Fe}{xxv} lines with transitions $1s^2 \,^1S_0 - 1snp\,^1P_1$ ($n=3$, 4, 5 etc.) and
associated \ion{Fe}{xxiv} satellites with transitions $1s^2 2p - 1s 2p nl$ ($n\geqslant
3$) which are well separated from each other, even when turbulent velocities of a few
hundred km~s$^{-1}$ are present. The purpose of this paper is to point out that these
lines are more suitable for the determination of temperatures from the hottest part of
solar flares, and may prove particularly useful for searching for nonthermal electrons
during flare impulsive stages.

\section{Spectral region 7.7 - 8.6 keV}

\begin{figure}
  \resizebox{\hsize}{!}{\includegraphics{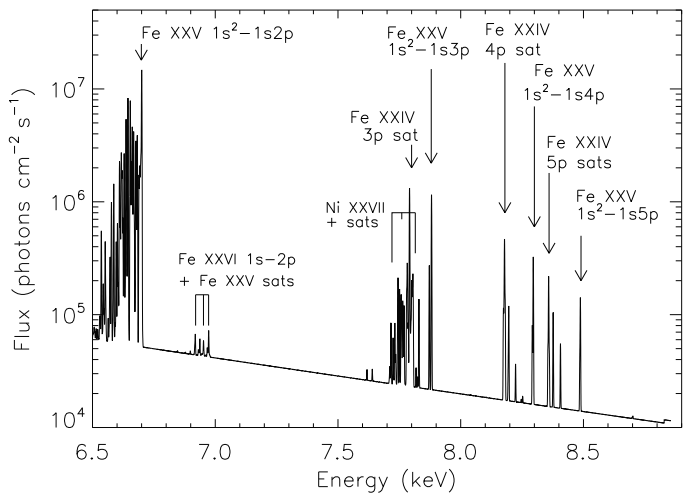}}
  \caption{Synthetic spectrum for solar flare plasma with $T_e = 20$~MK and volume emission measure
  $= 10^{49}$~cm$^{-3}$, plotted
  on a logarithmic flux scale, using the CHIANTI database and software (v. 5.2.1).
  Coronal abundances from \cite{fel92} are assumed. }
  \label{synth_T20MK}
\end{figure}

A selection of lines in the spectral range 7.7--8.6~keV is given in
Table~\ref{line_list}. The principal \ion{Fe}{xxv} lines with transitions $1s^2\,^1S_0 -
1snp\,^1P_1$ ($n=3$, 4, 5) are labelled $w_3$, $w_4$, $w_5$, and the much weaker
subsidiary lines with transitions $1s^2\,^1S_0 - 1snp\,^3P_1$ are labelled $y_3$, $y_4$,
$y_5$, following nomenclature of \cite{phi06a} for corresponding Si lines observed with
the RESIK spectrometer on {\it CORONAS-F}. Wavelengths (\AA) and photon energies (keV)
are listed, the data being from \cite{phi04} and the CHIANTI database and software
\citep{der97,lan06} (v. 5.2.1). (Note that from hereon, we specify lines by photon
energies in keV.) Figure~\ref{synth_T20MK} shows a synthetic spectrum (logarithmic flux
scale) in the range 6.5--8.8~keV, so includes the \ion{Fe}{xxv} $w$ line and associated
\ion{Fe}{xxiv} satellites. The spectrum was synthesized using CHIANTI (v. 5.2.1), and
assumes coronal excitation and ionization conditions and coronal abundances from
\cite{fel92} (these are negligibly different from those of the later work of
\cite{fel00}). The effective collisional excitation rates used in CHIANTI for the
\ion{Fe}{xxv} and \ion{Ni}{xxvii} lines are from the intermediate-coupling calculations
including auto-ionizing resonances of \cite{zha87}. The principal \ion{Fe}{xxv} lines
with transitions $1s^2\,^1S_0 - 1snp\,^1P_1$ ($n=3$, 4, 5) are important in the
7.7--8.5~keV region, as are groups of \ion{Fe}{xxiv} satellites with transitions $1s^2 \,
2p - 1s 2p nl$ ($n\geqslant 3$) covering small energy ranges. In Fig.~\ref{synth_T20MK}
they are labelled ``3p sat", ``4p sat", ``5p sat". The energy of each satellite group is
approximately 0.15--0.17~keV less than that of the corresponding ``parent" \ion{Fe}{xxv}
line with transition $1s^2\,^1S_0 - 1snp\,^1P_1$. Also present are \ion{Ni}{xxvii} lines,
including line $w$, and associated \ion{Ni}{xxvi} $1s^2 nl - 1s 2p nl$ ($n\geqslant 2$)
satellites, blending with the group of \ion{Fe}{xxiv} $1s^2 \, 2p - 1s 2p 3p$ satellites.
The temperature assumed in Figure~\ref{synth_T20MK} is $T_e = 20$~MK, a typical peak
flare temperature measured from the Fe lines near 6.7~keV. For higher temperatures, the
\ion{Fe}{xxiv} satellites (labelled 3p, 4p, 5p) are correspondingly less intense than the
corresponding \ion{Fe}{xxv} parent lines.

\begin{table*}
\caption{Principal lines in the 7.7--8.6 keV X-ray region. Note $a$: From CHIANTI, based
on solar coronal abundances, temperature 20~MK and volume emission measure
$10^{49}$~cm$^{-3}$.} \label{line_list} \centering
\begin{tabular}{l l l l l r}
\hline\hline
 Energy (keV)&Wavelength (\AA) &Ion & Line label &Transition  &Flux$^a$  \\

7.7573 & 1.598& Ni XXVI &     & $1s^2 2p\, ^2P_{1/2} - 1s 2p^2\, ^2D_{3/2}$ & 285 \\
7.7822 & 1.593& Fe XXIV &     & $1s^2 2p\,^2P_{3/2} - 1s 2p\, (^3P)\, 3p \,^4P_{5/2}$& 520  \\
7.7915 & 1.591& Fe XXIV &$j_3$& $1s^2 2p\,^2P_{3/2} - 1s2p \,(^3P)\, 3p \,^2D_{5/2}$ & 2810  \\
7.7924 & 1.591& Fe XXIV &     & $1s^2 2p\, ^2P_{1/2} - 1s 2p \,(^3P)\, 3p\, ^2D_{3/2}$ & 370 \\
7.8052 & 1.589& Ni XXVII &$w$& $1s^2\,^1S_0 - 1s2p\,^1P_1$  & 409 \\
7.8721 & 1.575& Fe XXV  &$y_3$& $1s^2\,^1S_0 - 1s3p\,^3P_1$ & 589 \\
7.8811 & 1.573& Fe XXV  &$w_3$& $1s^2\,^1S_0 - 1s3p\,^1P_1$ & 2650 \\
8.1785 & 1.516& Fe XXIV  &$j_4$& $1s^2 2p\,^2P_{3/2} - 1s2p \,(^3P)\, 4p \,^2D_{5/2}$ & 1030 \\
8.1812 & 1.516& Fe XXIV &     & $1s^2 2p\, ^2P_{1/2} - 1s 2p\, (^3P)\, 4p\, ^2D_{3/2}$ & 230\\
8.2917 & 1.495& Fe XXV  &$y_4$& $1s^2\,^1S_0 - 1s4p\,^3P_1$ & 151 \\
8.2956 & 1.495& Fe XXV  &$w_4$& $1s^2\,^1S_0 - 1s4p\,^1P_1$ & 724 \\
8.3554 & 1.484& Fe XXIV &     & $1s^2 2p\, ^2P_{3/2} - 1s 2p\, (^3P)\, 5p\, ^4P_{5/2}$ & 83 \\
8.3582 & 1.483& Fe XXIV  &$j_5$& $1s^2 2p\,^2P_{3/2} - 1s2p\, (^3P)\, 5p \,^2D_{5/2}$ & 473 \\
8.3610 & 1.483& Fe XXIV &     & $1s^2 2p \,^2P_{1/2} - 1s 2p\, (^3P)\, 5p\, ^2D_{3/2}$& 68 \\
8.3763 & 1.480& Fe XXIV &     & $1s^2 2s \,^2S_{1/2} - 1s 2s\, (^3S)\, 5p\, ^2P_{3/2}$ & 87 \\
8.3768 & 1.480& Fe XXIV &     & $1s^2 2s \,^2S_{1/2} - 1s 2s\, (^1S)\, 5p\, ^2P_{1/2}$ & 132 \\
8.4064 & 1.475& Fe XXIV &     & $1s^2 2s \,^2S_{1/2} - 1s 2s\, (^1S)\, 5p\, ^2P_{3/2}$ & 95 \\
8.4852 & 1.461& Fe XXV  &$y_5$& $1s^2\,^1S_0 - 1s5p\,^3P_1$ & 63 \\
8.4875 & 1.461& Fe XXV  &$w_5$& $1s^2\,^1S_0 - 1s5p\,^1P_1$ & 296 \\

\hline

\end{tabular}

\end{table*}

Each group of \ion{Fe}{xxiv} satellites in the 7.7--8.6~keV range is not only fairly
widely separated from its parent line, but is dominated by one particular satellite, with
transition $1s^2 \, 2p \,^2P_{3/2}- 1s 2p np\, ^2D_{5/2}$. The next most intense
satellite, with transition $1s^2 \, 2p \,^2P_{1/2}- 1s 2p np\, ^2D_{3/2}$, is at least a
factor 2 less intense. The grouping of such satellites in Si was noted by \cite{syl03}
and in laboratory spectra by \cite{fel74}.

\begin{figure}
  \resizebox{\hsize}{!}{\includegraphics{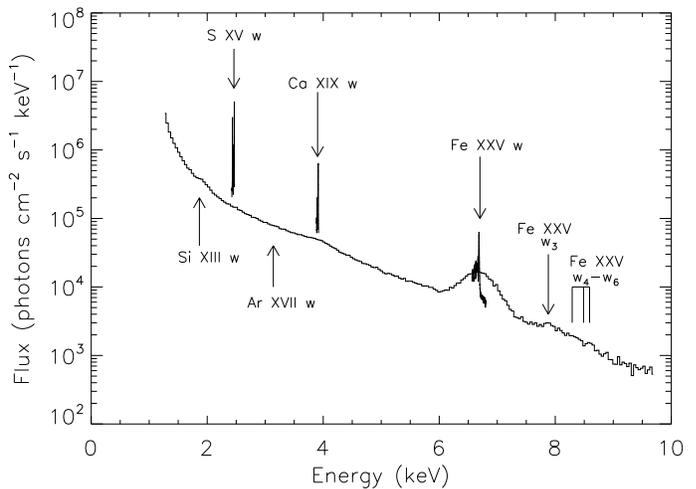}}
  \caption{Spectrum in the 1.2--9.7~keV range observed by the solar spectrometer {\it NEAR}-PIN during
  a flare on 1998 January~15, compared with spectra in narrow ranges around the \ion{S}{xv}, \ion{Ca}{xix},
  and \ion{Fe}{xxv} $w$ lines as observed with the Bragg Crystal Spectrometer on the {\it Yohkoh} spacecraft.
  The energies of other line groups are also indicated (only the \ion{Si}{xiii} line group is apparent). }
  \label{NEAR-PIN}
\end{figure}

There are very few solar flare observations of the spectral region 7.7--8.6~keV, but one,
reported by \cite{neu71}, was made during a large (optical class 2B) flare on 1969
February~27. The spectral range in question was covered by a crystal spectrometer using
LiF crystal ($2d = 4.027$~\AA). The spectrum shows a prominent line feature at $\sim
7.9$~keV ($\sim 1.56$~\AA, assuming a linear wavelength scale in the figure shown by
\cite{neu71}), and although identified as \ion{Ni}{xxvii} it is more likely to be the
\ion{Fe}{xxv} $1s^2\,^1S_0 - 1s3p\,^1P_1$ ($w_3$) line. Broad-band flare spectra in this
region have been observed by the PIN (solid-state detector) part of the X-ray
Spectrometer (XRS) on the {\it Near Earth Asteroid Rendezvous (NEAR)}--Shoemaker
spacecraft \citep{gol97} with spectral resolution of 600~eV at 5.9~keV. At this
resolution, the 6.7~keV line feature, which includes the \ion{Fe}{xxv} $w$ line and $n=2$
satellites, is unresolved as the ``Fe-line feature", while the higher-$n$ \ion{Fe}{xxv}
lines and associated \ion{Fe}{xxiv} satellites appear as a broad structure, with the
group of $n=3$ lines at $\sim 7.8$~keV barely resolved from the $n\geqslant 4$ lines at
8.2--8.5~keV. This can be seen from the spectrum shown in Figure~\ref{NEAR-PIN} taken
during a large flare on 1998 January~15. Superimposed on the {\it NEAR}-PIN spectrum,
which extends over the range 1.2--9.7~keV, are spectra from the {\it Yohkoh} Bragg
Crystal Spectrometer (BCS) which had narrow spectral windows centred on groups of lines
due to He-like S (\ion{S}{xv}), Ca (\ion{Ca}{xix}), and Fe (\ion{Fe}{xxv}). BCS
calibration factors were taken from data given by \cite{cul91} and incorporated in
standard IDL BCS analysis software. The continuum levels in the BCS and {\it NEAR-PIN}
spectra are within 40~\% (the BCS peak fluxes of the lines are much higher than the
corresponding bumps in the {\it NEAR-PIN} spectra owing to the higher spectral resolution
of the BCS).

Other broad-band spectral observations are available from the {\it Reuven Ramaty
High-Energy Solar Spectroscopic Imager} ({\it RHESSI}) with energy range $\sim 3$~keV to
17~MeV \citep{lin02}. For seven of its nine detectors in the 4--10~keV range the spectral
resolution is about 1~keV.  During the onset of a solar flare, attenuators are inserted
over the detectors so that the attenuator states during a large flare go from A0 (no
attenuator), through A1 (thin attenuators inserted) to A3 (thin and thick attenuators
inserted). The Fe-line feature is generally clearly visible in A1 and A3 spectra,
analysis of which has led to coronal Fe abundance determinations \citep{phi06b}. In A3
spectra, an emission line feature at $\sim 8$~keV is tentatively attributed to the
cluster of \ion{Fe}{xxv} lines, \ion{Fe}{xxiv} satellites, and \ion{Ni}{xxvii} lines to
form the ``Fe/Ni-line feature", though the nearby Fe-line (6.7~keV) feature and an
instrumental line feature at $\sim 10$~keV makes this line feature difficult to measure.

\section{Diagnostic potential of lines}

\subsection{Temperature estimates}

\begin{figure}
  \resizebox{\hsize}{!}{\includegraphics{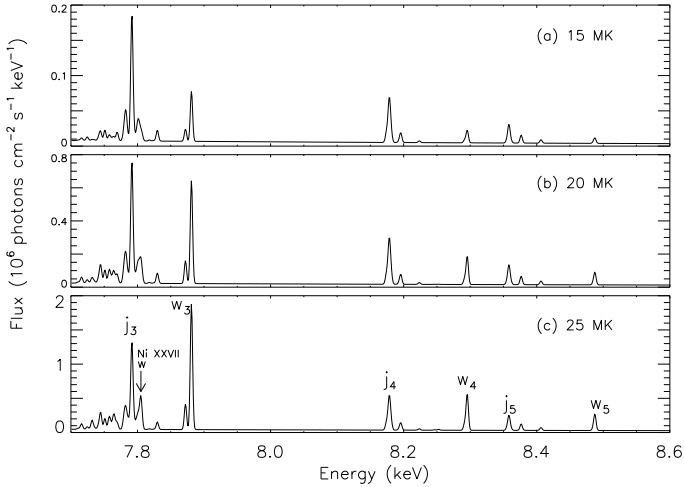}}
  \caption{Spectra in the 7.7--8.6~keV range synthesized from CHIANTI with solar coronal abundances,
  and spectral resolution $\Delta E \sim 2$~eV. The temperatures
  are indicated. Satellite line groups and parent lines are identified according to the text and
  Table~\ref{line_list}. }
  \label{3_synth_sp}
\end{figure}

Figure~\ref{3_synth_sp} shows spectra synthesized from CHIANTI in the 7.7--8.6~keV range
(linear flux scale) for $T_e = 15$~MK, 20~MK, and 25~MK, typical of moderate to very
intense solar flares. The spectral resolution chosen was equal to the thermal Doppler
width (FWHM) for ion temperature equal to $T_e$, i.e. $\Delta E = 3.0$~eV, 3.5~eV, and
3.9~eV respectively. The \ion{Fe}{xxiv} satellites with transitions $1s^2 \, 2p
\,^2P_{3/2}- 1s 2p np\, ^2D_{5/2}$ are labelled $j_3$, $j_4$, and $j_5$ for $n=3$, 4, 5,
by analogy with the \ion{Fe}{xxiv} satellite $j$ ($1s^2 \, 2p \,^2P_{3/2}- 1s 2p^2\,
^2D_{5/2}$ at 6.642~keV. Clearly, the fluxes of the \ion{Fe}{xxiv} satellites relative to
the corresponding \ion{Fe}{xxv} $w_3$, $w_4$, and $w_5$ lines decrease as $T_e$
increases.

Excitation of the satellite lines proceeds by dielectronic recombination of the He-like
stage, Fe$^{+24}$. This occurs first by dielectronic capture:

\begin{equation}
X^{+m}(1s^2) + e(nl) \leftrightarrows X^{+m-1}(1s 2p nl) \label{diel_cap}\end{equation}

\noindent (the left arrow indicating that the reverse, autoionization process may occur),
followed by stabilizing transitions:

\begin{equation}
X^{+m-1} (1s 2p nl) \rightarrow X^{+m-1}(1s 2p n'l') + h\nu
\end{equation}

\noindent where $n' < n$. If the configuration $1s 2p n'l'$ is $1s^2\,2p$, $h\nu =
h\nu_{\rm sat}$, the transition energy of one of the satellite lines in the $3p$, $4p$ or
$5p$ groups.

The satellite line flux $F_s$ is then proportional to the density of electrons $N_e$ and
Fe$^{+24}$ ions, to the dielectronic capture rate which by the principle of detailed
balance is proportional to $A_a/T_e^{3/2}$ ($A_a = $ the auto-ionization rate), and to a
branching ratio $A_r/(A_a + \Sigma A_r)$ indicating the relative probability of a
radiative transition occurring after the dielectronic capture process in
Eq.~\ref{diel_cap} (see \cite{gab72}). It is also proportional to ${\rm exp}\,(-E_s/k_B
T_e)$ where $E_s$ is the excitation energy of the doubly excited configuration $1s 2p nl$
above the Fe$^{+24}$ ground state $1s^2\, ^1S_0$ ($k_B$ is Boltzmann's constant). Thus

\begin{equation}
F_s = {\rm constant} \times N_e N({\rm Fe}^{+24}) \frac{A_r A_a}{A_a + \Sigma A_r} \frac
{{\rm exp} \, (-E_s / k_B T_e)} {T_e^{3/2}}.
\end{equation}

\noindent The flux of the corresponding \ion{Fe}{xxv} parent line, transition $1s^2 -
1snp$, which is mostly collisionally excited from the ground state, is

\begin{equation}
F_{par} = N_e N({\rm Fe}^{+24})\, C\, (1 + \alpha)
\end{equation}

\noindent where $C$ is the collisional rate coefficient, corrected by a factor $(1 +
\alpha)$ to allow for high-$n$ satellites blending with the parent line. $C$ is
approximately given by

\begin{equation}
C = \frac{8.63\times 10^{-6}\, \Upsilon }{T_e^{1/2}} {\rm exp}\, (-E_0/k_B T_e)
\end{equation}

\noindent where $\Upsilon$ is the effective collision strength of the transition, a
slowly varying function of $T_e$, and $E_0$ is the parent line excitation energy. Thus,
the flux ratio

\begin{equation}
\frac{F_s}{F_{par}} = {\rm constant} \times \frac{{\rm exp} \, [(E_0 - E_s)/k_B T_e]
}{\Upsilon \, T_e}
\end{equation}

\noindent where the constant includes all atomic constants. For typical flare
temperatures, $(E_0 - E_s)$ is rather small compared with $k_B T_e$ so the satellite
line-to-parent line flux ratio is close to $T_e^{-1}$. Incidentally, for these satellites
at solar flare densities ($N_e \sim 10^{11}$~cm$^{-3}$), inner-shell excitation of the
upper levels, e.g. from the $1s^2 2p \,^2P_{1/2}$, $^2P_{3/2}$ levels, does not
contribute significantly owing to the small population of these levels at solar flare
densities.

\begin{figure}
 \resizebox{\hsize}{!}{\includegraphics{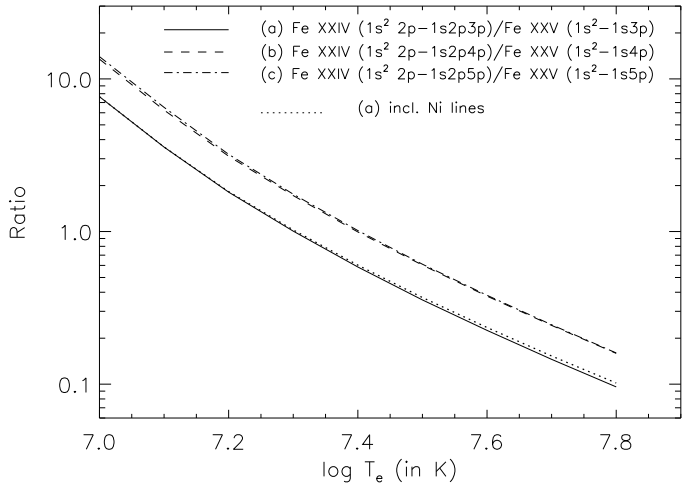}}
  \caption{Ratios of Fe XXIV satellite lines to Fe XXV parent lines, ($a$) $n=3$; ($b$) $n= 4$; ($c$) $n=5$
  satellites within 4~eV of the $j_n$ satellite line. For the $n=3$ satellites, the dotted line shows the
  contribution of Ni~XXVII and Ni~XXVI lines within 4~eV of the $j_3$ satellite. }
  \label{sat_parent_ratios}
\end{figure}

Figure~\ref{sat_parent_ratios} shows the line ratios of $3p$, $4p$, $5p$ satellite groups
to their corresponding parent lines plotted against ${\rm log}\,T_e$ (ratio scale
logarithmic). The \ion{Fe}{xxiv} satellites have been taken to be all those within 4~eV
of the $j_3$, $j_4$, $j_5$ satellite which is the one that dominates each satellite group
$3p$, $4p$, $5p$. Thus, measurements of the line ratios are assumed to be made with an
instrument having a spectral resolution $\sim 4$~eV. The ratios have again been estimated
from line fluxes given by CHIANTI. The \ion{Fe}{xxv} parent lines have been taken to
include both the allowed and intercombination lines $w_n$, $y_n$ (transitions
$1s^2\,^1S_0 - 1snp\,^1P_1$, $1s^2\,^1S_0 - 1snp\,^3P_1$),  but not unresolved satellites
with $n>5$ which are not included in CHIANTI. The fall-off of the ratio is not quite as
$T_e^{-1}$ owing to the slight temperature variation of the ${\rm exp} \, [(E_0 -
E_s)/k_B T_e]$ factor.

The \ion{Fe}{xxiv} $3p$ satellite group is blended with the \ion{Ni}{xxvii} lines and
associated \ion{Ni}{xxvi} $n=2$ satellites, and this needs to be taken account of in
Fig.~\ref{sat_parent_ratios}. The dotted line shows the ratio of the \ion{Fe}{xxiv} $3p$
satellite group with the Ni lines occurring within 4~eV of the \ion{Fe}{xxiv} $j_3$ line.
The Fe/Ni coronal abundance is 18 \citep{fel00}, so the contribution of the Ni lines is
always small, increasing with $T_e$.

\begin{figure}
 \resizebox{\hsize}{!}{\includegraphics{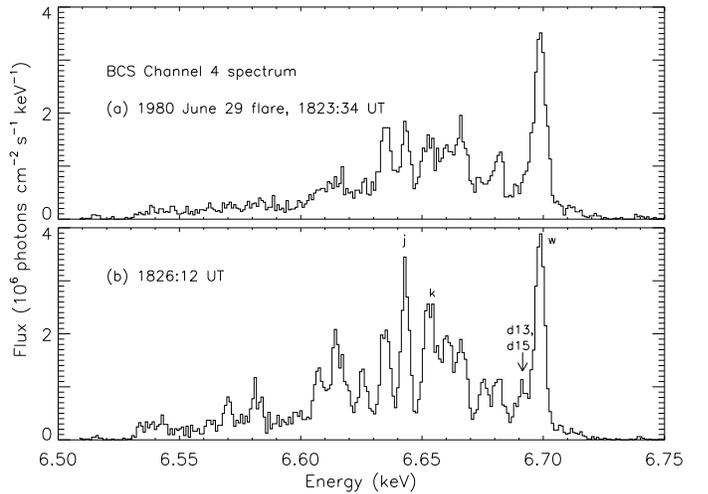}}
  \caption{Solar flare spectra from the {\it SMM} BCS during the limb flare on 1980 June 29: $a$
  at the flare impulsive stage,
  1823:34 UT. Integration time 12~s. Lines are broadened by a presumed turbulent mechanism, with
  measured width (FWHM) = 0.0096~keV (turbulent vely 170~km~s$^{-1}$). The $n=3$ satellite line feature
  made up of the $d13$, $d15$ lines
  is completely blended with the \ion{Fe}{xxv} $w$ line.  $b$
  after the flare impulsive stage, 1826:11 UT, integration time 12~s. Lines are broadened with measured
  FWHM = 0.00755~keV (turbulent vely 130~km~s$^{-1}$), sufficiently small that the $d13 + d15$ feature is
  now apparent. }
  \label{BCS_spectra}
\end{figure}

\begin{figure}
 \resizebox{\hsize}{!}{\includegraphics{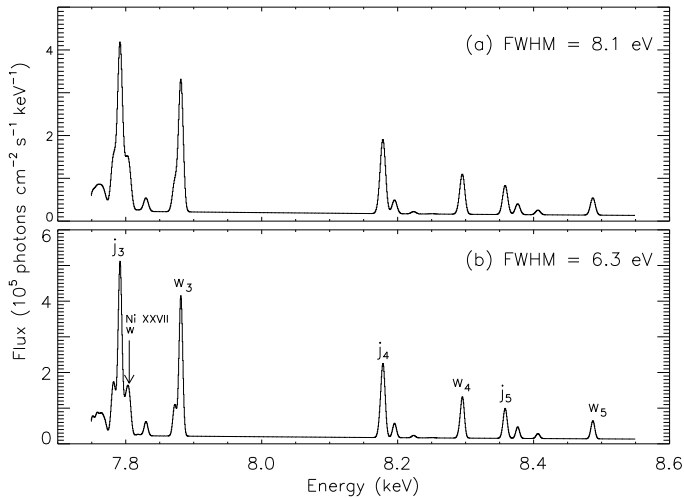}}
  \caption{Spectra in the 7.7--8.6~keV range synthesized with CHIANTI showing the principal \ion{Fe}{xxv} lines
  and  \ion{Fe}{xxiv} satellites. $a$ with FWHM 8.1~eV (equivalent to BCS spectrum in
  Fig.~\ref{BCS_spectra} $a$). $b$ with FWHM 6.3~eV (equivalent to BCS spectrum in
  Fig.~\ref{BCS_spectra} $b$). For both spectra, the principal \ion{Fe}{xxiv} satellites are well resolved from
  the \ion{Fe}{xxv} lines.  }
  \label{CHIANTI_broad_spectra}
\end{figure}

Although the satellites in the 7.7--8.6~keV range would appear to have little advantage
in the measurement of flare temperatures over that provided by the $n=2$ lines in the
6.7~keV range, the larger separation of the $3p$, $4p$, and $5p$ satellites from their
parent lines means that they are always much better resolved than the satellites near the
$w$ line at 6.70~keV. Figure~\ref{BCS_spectra} illustrates how at the flare impulsive
stage a spectrometer with good resolution -- in this case the BCS on {\it SMM} -- only
partly resolves the satellite line structure in the range 6.60--6.69~keV, including the
\ion{Fe}{xxiv} $j$ and $k$ lines. Fig.~\ref{BCS_spectra} ($a$) is the BCS spectrum during
the impulsive stage of a limb flare on 1980 June~29. Although there is no significant
blue-shifted component, the presence of turbulence means that the $j$ and $k$ lines are
sufficiently smeared with neighbouring satellites that temperature estimation must be
uncertain. A few minutes later (Fig.~\ref{BCS_spectra} ($b$)), the satellite lines are
well resolved and temperatures can be accurately estimated. However, the temperature
during the impulsive stage is of more importance for comparison with flare models
involving electron beam fluxes for which temperature, bulk and turbulent velocities are
predicted. Figure~\ref{CHIANTI_broad_spectra} shows CHIANTI simulations of the
7.7--8.6~keV region. In panel ($a$), the spectral resolution is 8~eV, equivalent to the
170~km~s$^{-1}$ turbulent velocity that is estimated from {\it SMM} BCS analysis software
for the spectrum in Fig.~\ref{BCS_spectra} ($a$); in panel ($b$), the spectral resolution
is 6~eV, equivalent to the smaller turbulent velocity (130~km~s$^{-1}$) in
Fig.~\ref{BCS_spectra} ($b$). In both cases, the $3p$, $4p$, and $5p$ satellite groups
are well resolved from their parent lines. In the case of Fig.~\ref{BCS_spectra} ($b$),
an estimate of the \ion{Ni}{xxvii} line contribution to the \ion{Fe}{xxiv} $3p$
satellites could be made by line-fitting techniques.

\subsection{Low-energy nonthermal electrons}

\begin{figure}
 \resizebox{\hsize}{!}{\includegraphics{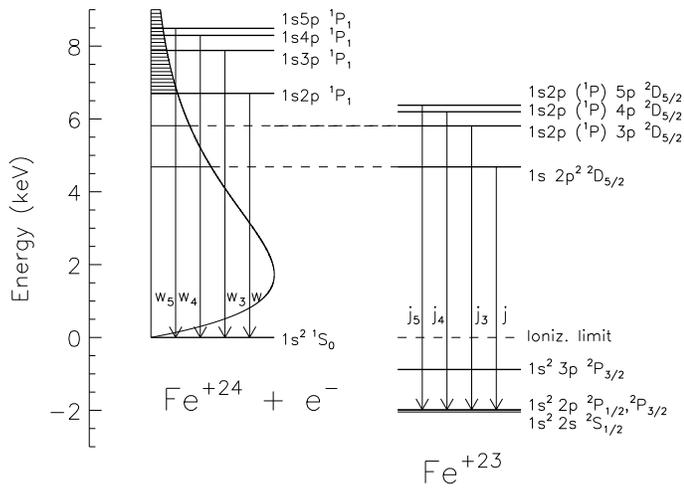}}
  \caption{Energy level diagram for ({\it left}) He-like Fe (Fe$^{+24}$ and ({\it right}) Li-like Fe
  (Fe$^{+23}$) $+ \,e$ to illustrate monoenergetic excitation of satellite lines. }
  \label{level_diagram}
\end{figure}

The method of \cite{gab79} to search for nonthermal electrons with energy of a few keV
relies on the comparison of temperature measurements using the $j/w$, $(d13 + d15)/w$,
and $(d13 + d15)/j$ line ratios. Figure~\ref{level_diagram} shows (right side) the energy
level diagram for Li-like Fe (Fe$^{+23}$) and (left side) He-like Fe (Fe$^{+24}$) plus a
free electron. Transitions forming the \ion{Fe}{xxv} $w$, $w_3$, $w_4$, and $w_5$ lines
are shown. The excitation energies of the upper levels of these lines are indicated on
the energy scale shown on the left-side of the diagram (energy of the ground state
$1s^2\,^1S_0$ equal to 0). Also shown is a Maxwell-Boltzmann distribution appropriate to
the free electron for $T_e = 20$~MK, a typical peak flare temperature. Excitation of any
of the \ion{Fe}{xxiv} satellites occurs by dielectronic capture of the free electron to
give the doubly excited upper levels of the satellites $j$, $j_3$, $j_4$, $j_5$,
$1s2p\,(^1P)\,np\,^2D_{5/2}$ ($n=2$ \dots 5). The free electron must have energies (to
within the auto-ionizing widths, $\lesssim 0.1$~eV) of 4.68~keV, 5.81~keV, 6.20~keV, and
6.38~keV for the excitation of the $n=2$ \dots 5 levels respectively. De-excitation
occurs from the $1s2p\,(^1P)np\,^2D_{5/2}$ levels either to the $1s^2 2p\, ^2P_{3/2}$
level or to the $1s^2 np\,^2P_{3/2}$ level. Taking the case $n=3$, the transition $1s^2
3p\,^2P_{3/2} - 1s2p\,(^1P)\,3p\,^2D_{5/2}$ corresponds to the $d13$ satellite at
6.694~keV, the transition $1s^2 2p\,^2P_{3/2} - 1s2p\,(^1P)3p\,^2D_{5/2}$ to the $j_3$
satellite at 7.792~keV. Note that the $d15$ line (transition $1s^2 3p\,^2P_{1/2} -
1s2p\,(^1P)\,3p\,^2D_{3/2}$) is blended with $d13$, and has a flux of $0.73 \times d13$.

Formation of the doubly excited states by dielectronic capture ``samples" the
Maxwell-Boltzmann distribution at energies 4.68~keV, 5.81~keV, 6.20~keV, and 6.38~keV.
The method of \cite{gab79} was to estimate $T_e$ from the $j/w$ ratio ($T_1$) or from the
$(d13 + d15)/w$ ratio ($T_2$) and compare with the temperature from the $(d13 + d15)/j$
ratio ($T_3$). If the flaring plasma can be assumed to be isothermal but if $T_1$ and
$T_2$ are significantly higher than $T_3$, nonthermal electrons may be implied at higher
energies, e.g. $\gtrsim 10$~keV, as are commonly deduced from hard X-ray spectra during
the flare impulsive stage. If the number of nonthermal electrons is insignificant at
energies less than the threshold energy of the $w$ line excitation (6.702~keV),
temperature $T_3$ is the true flare temperature (again, on the isothermal assumption)
since it is measured by the numbers of electrons at energies 4.68~keV and 5.81~keV (upper
levels of $j$ and $d13$).

In practice, it has been found that, at the flare impulsive stage when nonthermal
electrons are expected to be present, line broadening due to turbulence is such that the
satellite feature $d13+d15$ is blended with line $w$ (see the {\it SMM} BCS spectrum,
Fig.~\ref{BCS_spectra} ($a$)). However, some success in applying the \cite{gab79} method
has been achieved using synthetic spectra to match observed spectra during flare
impulsive stages by \cite{see87}. These authors used nonthermal electron distributions of
the form

\begin{equation}
F(E) =\frac { E^{n/2} \, {\rm exp} \,(-E/k_B T)}{\Gamma(n/2 + 1)\,(k_B T)^{n/2 + 1} }
\end{equation}

\noindent where $\Gamma$ is the gamma function. When the parameter $n=1$, $F(E)$ reduces
to the Maxwell-Boltzmann distribution with $T = T_e$. The flux ratios $j/w$, $(d13 +
d15)/w$, and $(d13 + d15)/j$ can be related to expressions that are functions of $n$ and
$T$. The \ion{Fe}{xxv} and \ion{Fe}{xxiv} lines at $\sim 6.7$~keV were observed  with the
SOLFLEX instrument on {\it P78-1} during three flares, with hard X-ray bursts at the
impulsive stages independently indicating the presence of nonthermal electrons. Very high
values of $n$ ($>15$) were derived in the impulsive stages which decreased to $n\sim 1$
during the flare thermal phases. \cite{see87} attribute these to departures from
Maxwell-Boltzmann distributions and not to non-isothermal (differential emission measure)
effects since the $(d13 + d15)$ feature is expected to be excited by more energetic
electrons than the $j$ satellite but during the flare impulsive stage the $(d13 + d15)$
feature is actually enhanced relative to $j$.

Clearly, observation of the \ion{Fe}{xxiv} $3p$, $4p$, and $5p$ satellite groups in the
7.7--8.6~keV region have the distinct advantage of not suffering from blending with their
parent lines, even during flare impulsive stages, as well as the fact that for typical
flare temperatures (15--20~MK) the satellites have comparable fluxes to the parent lines.
This is illustrated by the {\it SMM} BCS spectra for a limb flare shown in
Figure~\ref{BCS_spectra} ($a$), in which the weak $d13+d15$ line feature is completely
blended with the \ion{Fe}{xxv} line $w$, but for the same amount of line broadening for
the 7.7--8.6~keV range all the satellite groups show as strong features well separated
from the parent lines. Note that for disk flares, a short-wavelength component attributed
to rising plasma would give rise to increased blending with the 6.7~keV lines at the
resolution of the BCS. Even a spectrometer with perfect spectral resolution would have
blending problems for the 6.7~keV line group because of the presence of
turbulence-broadened lines and short-wavelength components.

Thus the $3p$, $4p$, and $5p$ satellite line groups are available for searching for
nonthermal electrons since the upper levels of the satellites in each group are excited
by mono-energetic electrons. In the case of the $j_3$, $j_4$, and $j_5$ satellites (see
Figure~\ref{level_diagram}), the upper levels are $1s2p\,(^1P)\,np\,^2D_{5/2}$, $n=3$, 4,
5, with excitation energies 5.81~keV, 6.20~keV, 6.38~keV respectively. The method of
\cite{gab79} can therefore be applied to these three satellites which are in practice
always strong relative to the \ion{Fe}{xxv} parent lines and well resolved, even during
flare impulsive stages.

\section{Future observations}

The 7.7--8.6~keV range has not been selected for solar flare spectrometers in the past
presumably because the \ion{Fe}{xxv} lines are appreciably weaker than the very strong
\ion{Fe}{xxv} $w$ line at 6.70~keV -- CHIANTI simulations indicate that for a temperature
20~MK, with the $w$ line having unit total flux (photon units), the $w_3$, $w_4$, and
$w_5$ lines have total fluxes of 0.08, 0.02, 0.01.  However, as is shown here, the
diagnostic potential of this range is considerable, and only modest spectral resolution
is needed to obtain line flux ratios to obtain the temperatures of the hottest regions of
flares and information about low-energy nonthermal electrons. Spectra in this region
could also be of importance for diagnosis of tokamak plasmas.

The range is available to crystal spectrometers, particularly those with bent geometries
such as were used for the {\it SMM} Bent Crystal Spectrometer and the {\it Yohkoh} Bragg
Crystal Spectrometer since their effective areas are much higher than equivalent scanning
flat crystal spectrometers. Possible diffracting crystals include germanium (Ge 422),
with $2d$ spacing equal to 2.31~\AA, though its integrated reflectivity is a factor 2
lower in the required range than at 6.70~keV, and lithium fluoride, LiF (420) with $2d =
1.80$~\AA. The development of X-ray microcalorimeters (instruments sensing tiny heat
pulses created by incident X-ray photons) offers some possibilities for observing this
range. The recent X-ray Spectrometer (XRS) on the {\it Suzaku} spacecraft \citep{kel07}
consisted of an array of microcalorimeters, and was designed to observe non-solar sources
in the spectral range 0.3--12~keV with a resolution of only 6~eV, easily enough to
resolve the \ion{Fe}{xxiv} satellite line structure at 6.55--6.70~keV. Although it did
not successfully make any observations owing to an early loss of liquid helium cryogen
needed to maintain a temperature of 60~mK, the designed spectral resolution was
apparently achieved \citep{kel07}. An instrument like this, capable of observing the
7.7--8.6~keV region with a time resolution high enough to follow the hard X-ray bursts at
the onset of solar flares, would be ideal for the temperature determination of the
hottest parts of the flare plasma and detection of nonthermal electrons.

\begin{acknowledgements}
I am grateful for helpful discussions with Janusz Sylwester (Polish Academy of Sciences,
Wroc{\l}aw, Poland) and Brian Dennis (NASA Goddard Space Flight Center), and for
permission to use {\it NEAR-Shoemaker} PIN data granted by Richard Starr (GSFC). CHIANTI
is a collaborative project involving Naval Research Laboratory (USA), University College
London and Cambridge University (UK), George Mason University (USA), and Florence
University (Italy).

\end{acknowledgements}

\end{document}